\documentclass[%
 reprint,
 amsmath,amssymb,
 aps,
]{revtex4-1}

\usepackage{enumitem}
\usepackage{amsmath}
\usepackage{graphicx}% Include figure files
\usepackage{dcolumn}% Align table columns on decimal point
\usepackage{bm}% bold math

\usepackage{color}
\definecolor{darkgreen}{rgb}{0.0, 0.5, 0.0}

\newcommand{\IMP}[1]{\textcolor{black}{#1}}%{\bf #1}}

\newcommand{\DEM}{D_\mathrm{ex}(f; t - t_i)}
\newcommand{\DEMshort}{D_\mathrm{ex}}
\newcommand{\aDEM}{{\tilde D}_\mathrm{ex}(f; t - t_i)}
\newcommand{\aDEMshort}{{\tilde D}_\mathrm{ex}}

\begin{document}

\title{The sounds of failure: passive acoustic measurements of excited vibrational modes}

\author{Theodore A. Brzinski III}
 \email{tbrzinski@haverford.edu}
 \altaffiliation[Now at ]{Department of Physics and Astronomy, Haverford College.}
\author{Karen E. Daniels}%
\affiliation{Department of Physics, NC State University}%

\begin{abstract}
Granular materials can fail through spontaneous events like earthquakes or brittle fracture.
However, measurements and analytic models which forecast failure in this class of materials, while of both fundamental and practical interest, remain elusive.
Materials including numerical packings of spheres, colloidal glasses, and granular materials have been known to develop an excess of low-frequency vibrational modes as the confining pressure is reduced. 
Here, we report experiments on sheared granular materials in which we monitor the evolving density of excited modes via passive monitoring of acoustic emissions.
We observe a broadening of the distribution of excited modes coincident with both bulk and local plasticity, and evolution in the shape of the distribution before and after bulk failure.
These results provide a new interpretation of the changing state of the material on its approach to stick-slip failure.
 \end{abstract}

\maketitle

The framework provided by the {jamming transition}  \cite{jamming,Liu2010,VanHecke2010} has highlighted the extent to which granular and other amorphous systems share certain properties: spatially and dynamically heterogeneous response to stress, structural disorder, and inhomogeneous force transmission. 
While the onset of rigidity in jammed materials shares features with standard second-order phase transitions, jamming differs from other such transitions by its lack of a diverging structural length-scale.
Although jammed systems are not necessarily thermal, it has been observed that the density of vibrational modes $D(\omega)$ remains an important descriptor of the state of the system.
In particular, an excess of low-frequency modes develops on the approach to the jamming transition \cite{OHern2003,Wyart2005} and the onset of plasticity \cite{Tanguy2010}.
Indeed, these excess low-frequency modes have been observed in experiments in colloidal systems  \cite{Ghosh2010,Chen2010} as well as granular materials  \cite{Brito2010,Owens2013}.
    
As such, both experiments and simulations tantalizingly suggest that information about the rigidity of a system might be encoded within $D(\omega;t)$ as the system evolves under external loading (thus the notation to denote its values at a specific time $t$).
$D(\omega;t)$ is a particularly attractive metric since the passive recording of acoustic emissions provides a non-invasive method of reporting changes in the state of the system and does not require visual access to the system.
For example, embedded sensors have long been used for non-destructive evaluation of engineered structures   \cite{Dunegan1968,Nair2010}, and have also successfully identified precursors in volcanic systems   \cite{Paparo2002}.

One practical method for measuring $D(\omega;t)$ has been to take advantage of the relationship between the particle velocity autocorrelation function and the density of vibrational modes  \cite{Dickey1969,Keyes1997}.
Recent experiments on a quasi-2D granular packing have used this relationship to establish a connection between acoustic modes and the jamming framework \cite{Owens2013}.
The procedure is to measure the velocity autocorrelation function
\begin{equation}
C_v \left(\tau;t\right) \equiv \frac{\sum v_{k}\left(t+\tau\right)v_{k}\left(t\right)}{\sum v_{k}\left(t\right)v_{k}\left(t\right)},
\label{eq:VAC}
\end{equation}
which is a function of both time ($t$) and lagtime ($\tau$). Here, $v_{k}(t)$ is the velocity timeseries measured using the $k$th of many particle-scale piezoelectric sensors; the sums over $k$ cover all sensors in the system. The density of vibrational modes is then given by  \cite{Owens2013} 
\begin{equation}
D\left(\omega;t\right)=\int_{0}^{\infty} C_{v}\left(\tau;t\right)\cos{\left(2\pi\omega\tau\right)} \, d\tau.
\label{eq:Dw}
\end{equation}
This approach succeeds even for measurements over a small subset of the particles, recovering the expected Debye scaling for crystalline granular materials as well as the expected excess of low-$\omega$ modes in both amorphous and crystalline systems as the confining pressure was reduced. 

Our experiments are inspired by prior work on slowly-loaded granular materials, from which it is known that (1) particle-scale rearrangements both precede and follow failure events  \cite{Nasuno1997}, and (2) acoustic emissions and microslips show an exponential increase in their rate of occurrence leading up to a failure event  \cite{Garcimartin1997,Johnson2013} and have been shown to encode information about the internal strength of the material~\cite{BRL2018}.
Here, we measure the acoustic emissions during the lead-up to failure, and associate changes in the observed $D(\omega;t)$ with the approach to failure.
In doing so, we provide a new means of acoustic monitoring.
Unlike spectral power measurements, which capture the distribution of acoustic \emph{power} among modes of different frequencies, the approach we introduce is effectively a measurement of the \emph{number} of modes which are excited, regardless of the excitation amplitude.
To differentiate from the actual density of vibrational modes, we denote this measurement as the density of excited modes, \(\DEMshort(f;t)\), where \(f\) replaces \(\omega\) as frequency.
As far as we know, there are no theoretical expectations for the behavior of \(\DEMshort\); what follows is an empirical exploration.

\begin{figure}
\centering\includegraphics[width=0.75\linewidth]{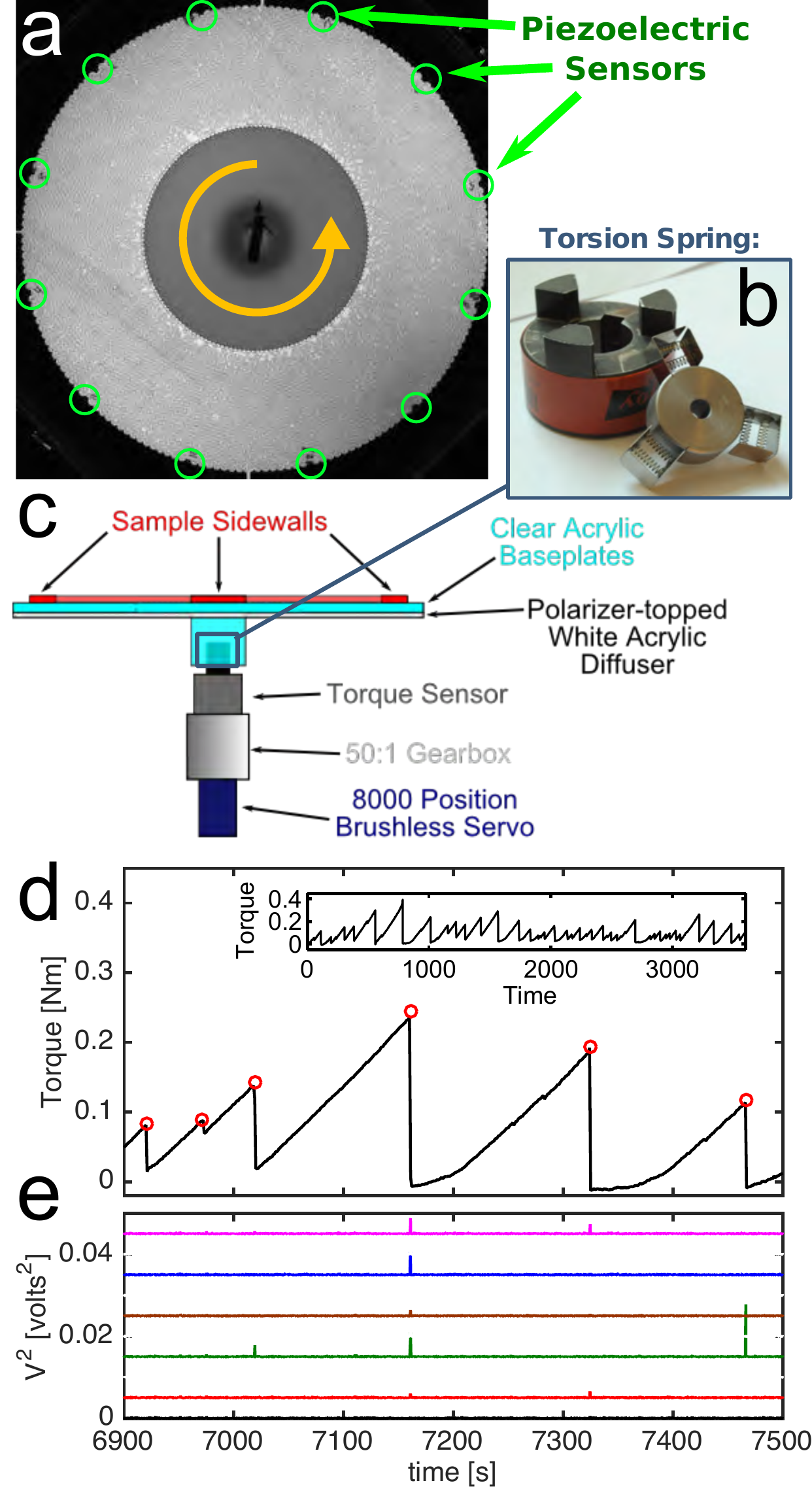}
\caption{\label{fig:apparatus}
{(a)} Top view of the annular shear cell, with piezoelectric sensors and moving (inner) wall. 
{(b)} Driveshaft coupling with Hookean torsion spring. 
{(c)} Schematic side view illustrating the details of the drive system. 
{(d)} Sample torque data over the course of $10$~minutes and (inset) $1$~hour.
Red circles identify slips.
{(e)} Sample measurements of the voltage squared, $V^2(t)$, measured for 5 piezoelectric sensors, vertically offset and plotted on the same time axis as {(d)}.
}
\end{figure}

Our experiment, depicted in Fig.~\ref{fig:apparatus}{a-c}, comprises an annulus with an outer wall of diameter $66.75$~cm and an inner wall of diameter $30.5$~cm.
The system is filled with a single layer of approximately $8000$ grains.
The grains are a $60$:$40$ mixture of $5.6$~mm circular and $4.9$ by $6.9$~mm elliptical disks to prevent crystallization.
All grains are milled from PhotoStress Plus PS-3 polymer from the Vishay Measurements Group with a bulk elastic modulus of $0.21$~GPa.
The granular material is sheared at a rate of 1 rotation per hour via a torsion spring (pictured in Fig.~\ref{fig:apparatus}{b}) with a stiffness of $0.85$~Nm/radian and maximum compression of $26^{\circ}$, corresponding to a torque of $0.39$~Nm.
Twelve (12) piezoelectric sensors, which are embedded in the outer wall, produce a voltage proportional to any compressive force they experience, thus registering a measurement of the acoustic emissions of the granular material.
During an experiment, the driving torque (Fig.~\ref{fig:apparatus}{d}) and acoustic emissions (Fig.~\ref{fig:apparatus}{e}) of the system are continuously measured via a torque sensor (Cooper Instruments) and acoustic sensors (detailed in  \cite{Owens2011}).

This driving produces intermittent stick-slip failure events, apparent as the sawtooth features in Fig.~\ref{fig:apparatus}{d}.
The loading (stick) phase corresponds to the compression of the torsional spring, and when the applied torque surpasses the strength of the granular material, failure (slip) occurs.
Empirically, after an initial transient, the frequency of slips remains close to $1$~slip/minute.
The data presented here were all collected in this steady state over the course of $23$~hours.
The full dataset of $1165$ slip events are aperiodic and span a broad range of torque and time scales (see the inset in Fig.~\ref{fig:apparatus}{d}), indicating substantial heterogeneity in the material strength and degree of deformation.
Despite the spatiotemporal heterogeneity, the slip durations exhibit a relatively narrow distribution with a mean of $0.65\pm0.14$~s.
This slipping timescale is well-separated from the inter-slip (quiescent) timescale: more than $80\%$ of slips, are proceeded and followed by quiescent periods of $30$~seconds or more.
We focus our analysis on the subset of these $887$ slips to isolate the effects of individual events.

An illustration of the typical intermittency of the acoustic emissions is provided in Fig.~\ref{fig:apparatus}{e}; each trace is the power from one piezoelectric sensor.
We observe that the largest emissions always coincide with slips.
Importantly, the converse is not the case: not every slip produces a voltage spike in every sensor.
This behavior arises because the force chains cause spatial heterogeneities in acoustic transmission \cite{Owens2011}.
The typical RMS voltage during quiescent periods (without large torque drops) is $2.0$~mV, the noise floor of our data acquisition hardware is $1.34$~mV, and emission events can produce spikes as much as $3$ orders of magnitude higher.
In this paper, we investigate the low-amplitude emissions during the largely quiescent periods between these slips.

\begin{figure}
\centering\includegraphics[width=\linewidth]{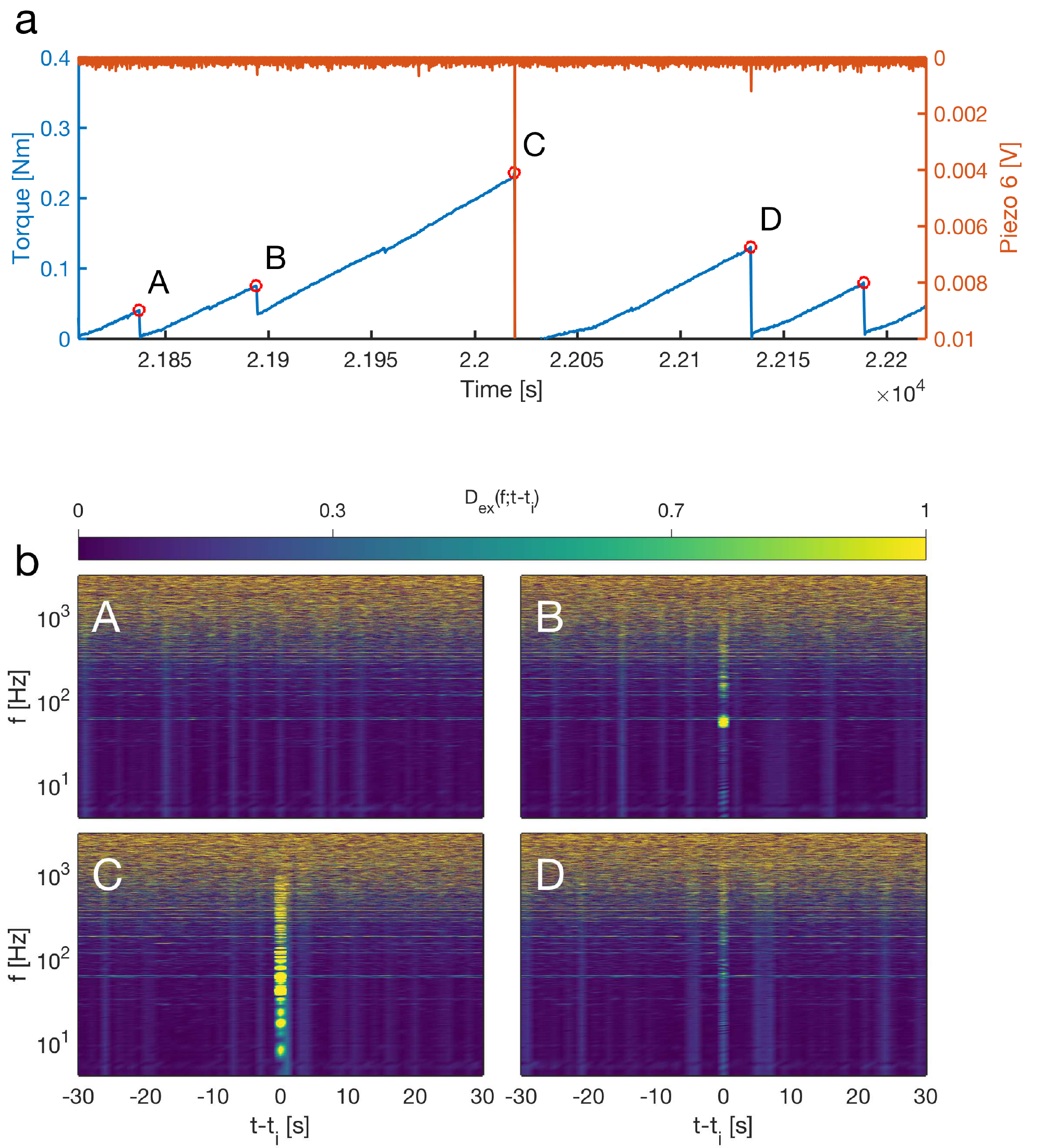}
\caption{\label{fig:SmplSpcts} 
{(a)} $400$~s of torque (blue) and voltage (red) measurements from one piezoelectric sensor. 5 slips are labelled with red circles. 
{(b)} Sample modograms showing the density of excited modes $\DEM$ for an interval of $\pm 30$~s centered on each of the four labled slips (A-D).}
\end{figure}

We analyze the evolution of the density of excited states, $\DEM$, by the following procedure:
 For each sensor and slip event occurring at time \(t_i\), we divide the voltage timeseries into 61 $1$~s intervals centered around \(t_i\); we integrate the voltage over each interval to obtain the sensor velocities in arbitrary units; we calculate $D_\mathrm{ex}(f)$ via Eqs.~\ref{eq:VAC}~\&~\ref{eq:Dw} \cite{Dickey1969,Keyes1997} (see Supp. Mat.); we plot this quantity as a function of frequency \(f\) and \(t-t_i\) in \emph{modograms}.
Sample modograms for four events are shown in Fig.~\ref{fig:SmplSpcts}{b}.
One prominent feature is that a broad range of modes is excited during each slip  ($t = t_i$).
However, similar increases in excited modes are also observed at times before and after the slips; these features are visible as bright vertical bands in Fig.~\ref{fig:SmplSpcts}{b}.
We also observe that the overall density of excited modes appears to be relatively flat over nearly 3 decades (Hz to kHz), rising only at the highest frequencies in this range.
Finally, we observe several persistent frequencies which appear as horizontal lines.
While the $60$~Hz peak is of electronic origin, the others are likely due to acoustic noise such as from the drive system or building noise.
These noise peaks will be filtered out in the later stages of analysis. 

We begin by focusing on the vertical bands visible in Fig.~\ref{fig:SmplSpcts}{b}.
Each of these bands represents a time at which a single piezoelectric sensor detected an increase in the number of excited vibrational modes over a broad range of frequencies.
Some of these vertical bands correspond to global slip events ($t=t_i$), but most are detected due to local rearrangements which happened to occur close to a particular sensor.
The local nature of these detections is reinforced by the observation that modograms from different sensors do not all record increases (as also seen in Fig.~\ref{fig:apparatus}{e}).
The ability to measure the $\DEM$ from either low- or high-amplitude slip events is crucially-important to this method. 
\IMP{While the low-amplitude events are too small to cause a global slip event, some of them are nonetheless large enough to be detected as they travel through the granular material and thereby leave a record of the state of the material. As we shall see below, the information they transmit reveals the changing state of the material.}

\begin{figure}
\centering\includegraphics[width=0.8\linewidth]{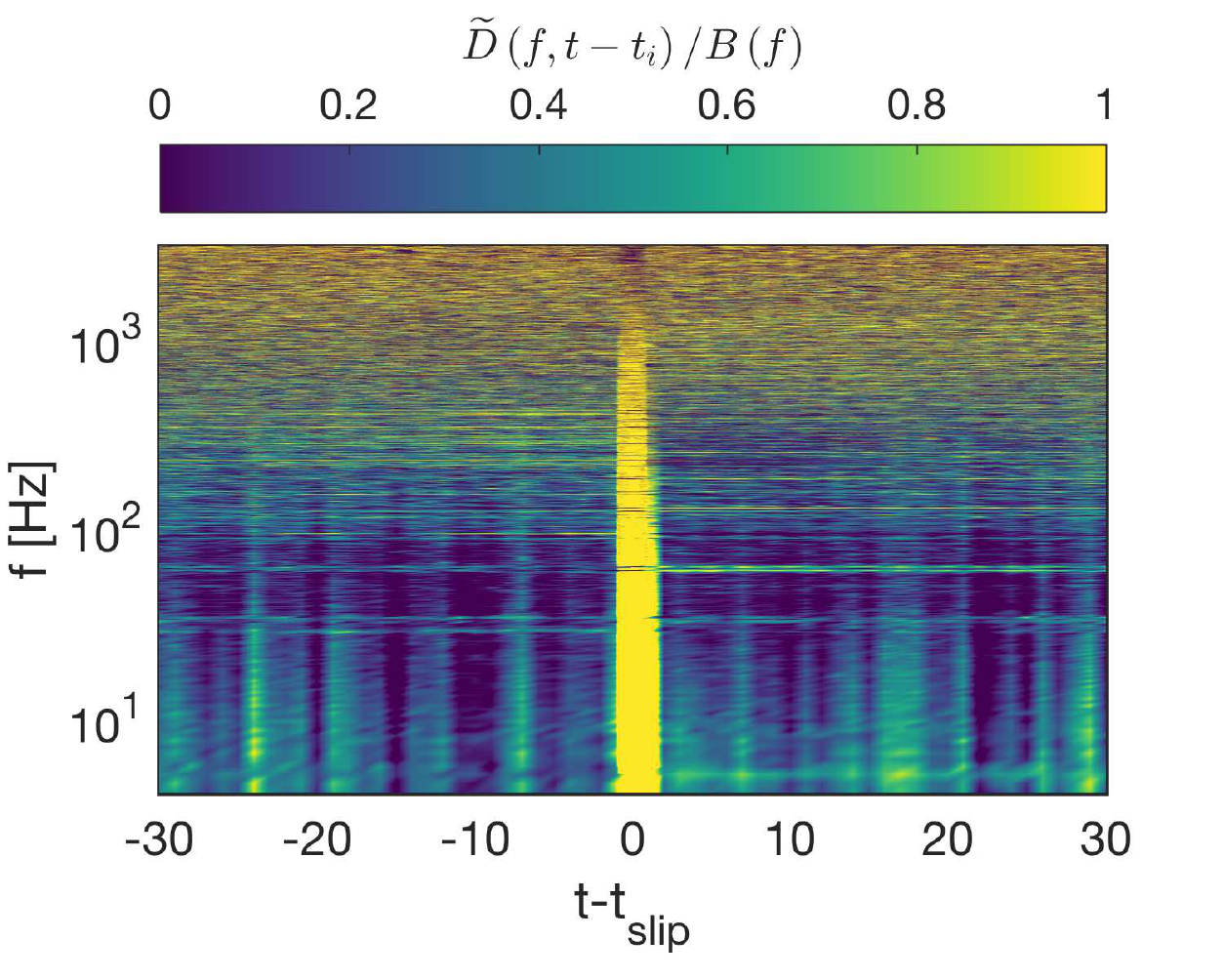}
\caption{\label{fig:mnDoMs}
Average modograms $\aDEM$, taken over 887 slips and 12 sensors and normalized by the sensor- and time-averaged density of excited states, $B(f)$.
}
\end{figure}

In order to reduce the impact of noise in the density of excited modes, we construct the average modogram  $\aDEM \equiv  \langle \DEM \rangle_{i,k}$, where $i$ is an index over the 887  detected slips separated by at least 30~s from the adjacent slips, and $k$ is an index over the 12 sensors.
To highlight the relative changes this quantity exhibits in response to failure, and to supress persistent electronic/physical resonances, we normalize $\aDEMshort$ by the sensor- and time- averaged density of excited states, $B(f) = \langle \aDEM \rangle_{t\ne t_i}$ to obtain the rescaled modogram shown in Fig.~\ref{fig:mnDoMs}.%{  b}. 

This normalized average modogram exhibits features similar to the fluctuations observed in Fig.~\ref{fig:SmplSpcts}, but now the vertical streaks result from the average behavior over {\it many} slips and sensors.
The remaining heterogeneity indicates that 887 slips were an insufficient quantity of data to eliminate the temporal heterogeneity associated with localized plasticity.
Even within this noisy signal, however, there emerge clear differences between the pre- and post-slip portions of this modogram.

\begin{figure}
\centering\includegraphics[width=\linewidth]{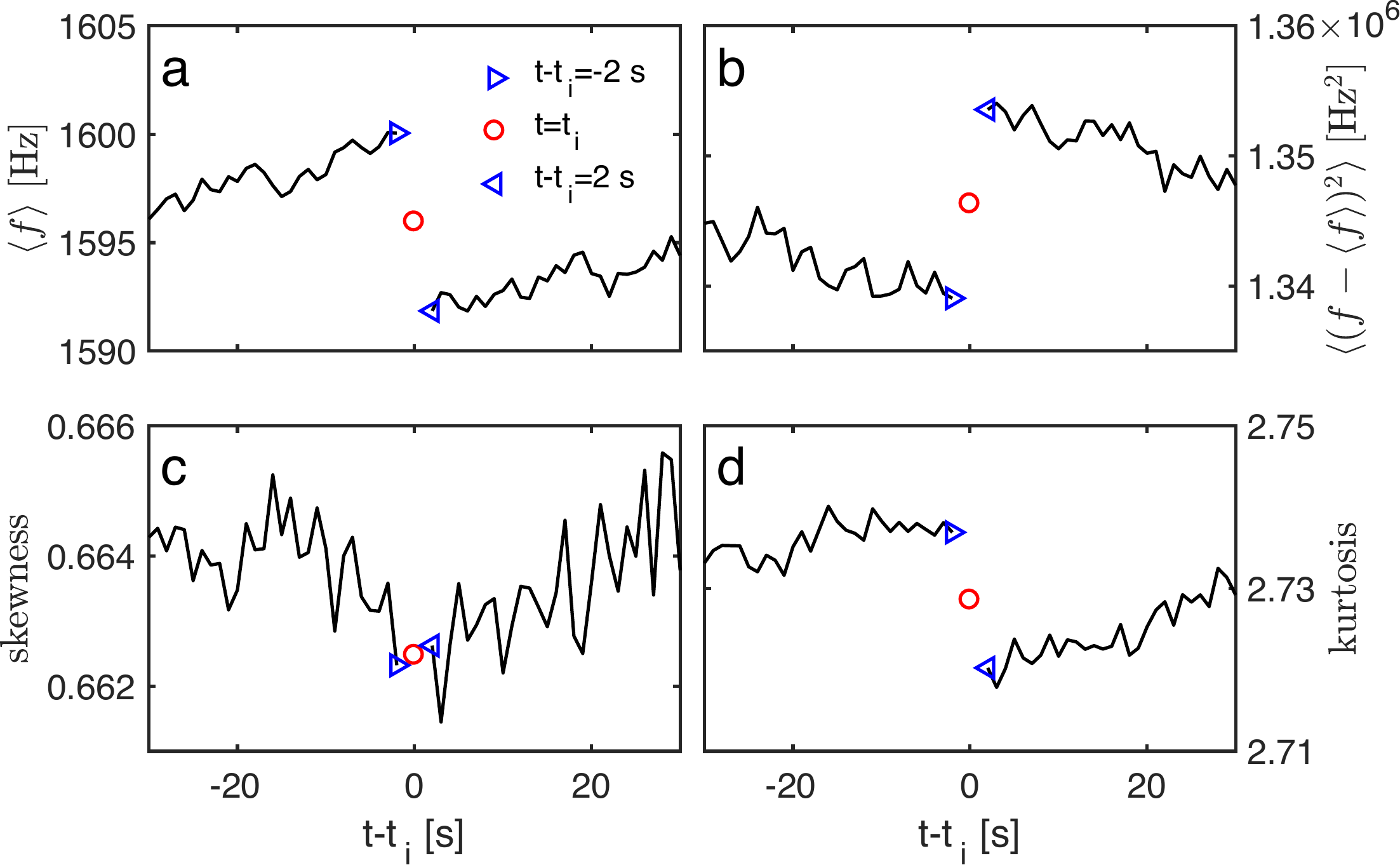}
\caption{\label{fig:MeanFreq} 
The four moments of $\aDEM$, calculated at fixed times pre- and post-slip: {(a)}  mean frequency, {(b)}  variance, {(c)}  skewness, and {(d)}  kurtosis. The red `$\circ$' indicates the time of the slip (datapoint omitted for clarity). The right- and left-pointing cyan triangles indicate the values immediately before and after the slip.}
\end{figure}

To characterize the changes in the slip- and sensor-averaged $\aDEMshort$, we calculate the first four moments of that quantity.
These are best considered as empirical shape parameters, since there is no prediction for the shape of $\aDEMshort$, and our dynamic range may capture only a portion of the distribution.
Nonetheless, a clear signal is evident in these quantities (Fig.~\ref{fig:MeanFreq}).
We find that the mean frequency ({a}) gradually grows during the pre-slip phase, and then suddenly decreases from $1.60$ to $1.59$~kHz (less than 1\%) in response to slips. 
This effect is accompanied by a small increase in the variance ({b})
\IMP{If our dynamic range captures most of the excited modes, these changes are consistent with a broadening of $\aDEMshort(f)$ in response to failure, with excited modes arising at lower-frequencies.} 
Since the internal stress in the granular material is lower after a slip (see Fig.~\ref{fig:apparatus}{d}), these observations are consistent with the observations of failure of the force chain network~\cite{veje1999} and emergence of excess low-frequency modes~\cite{Owens2013} at smaller confining stresses.

We also observe similarly clear, but still small, signals in the higher central moments of $\aDEMshort$: a weak minimum in skewness and a 1\% drop in the kurtosis (panels {c}-{d}, respectively). 
We find a skewness of $0.66$, which means low-frequency modes are more common than high-frequency modes. The kurtosis close to $3$ indicates $\aDEMshort$ is neither particularly heavy- nor weak-tailed.

\begin{figure}
\includegraphics[width=0.85\linewidth]{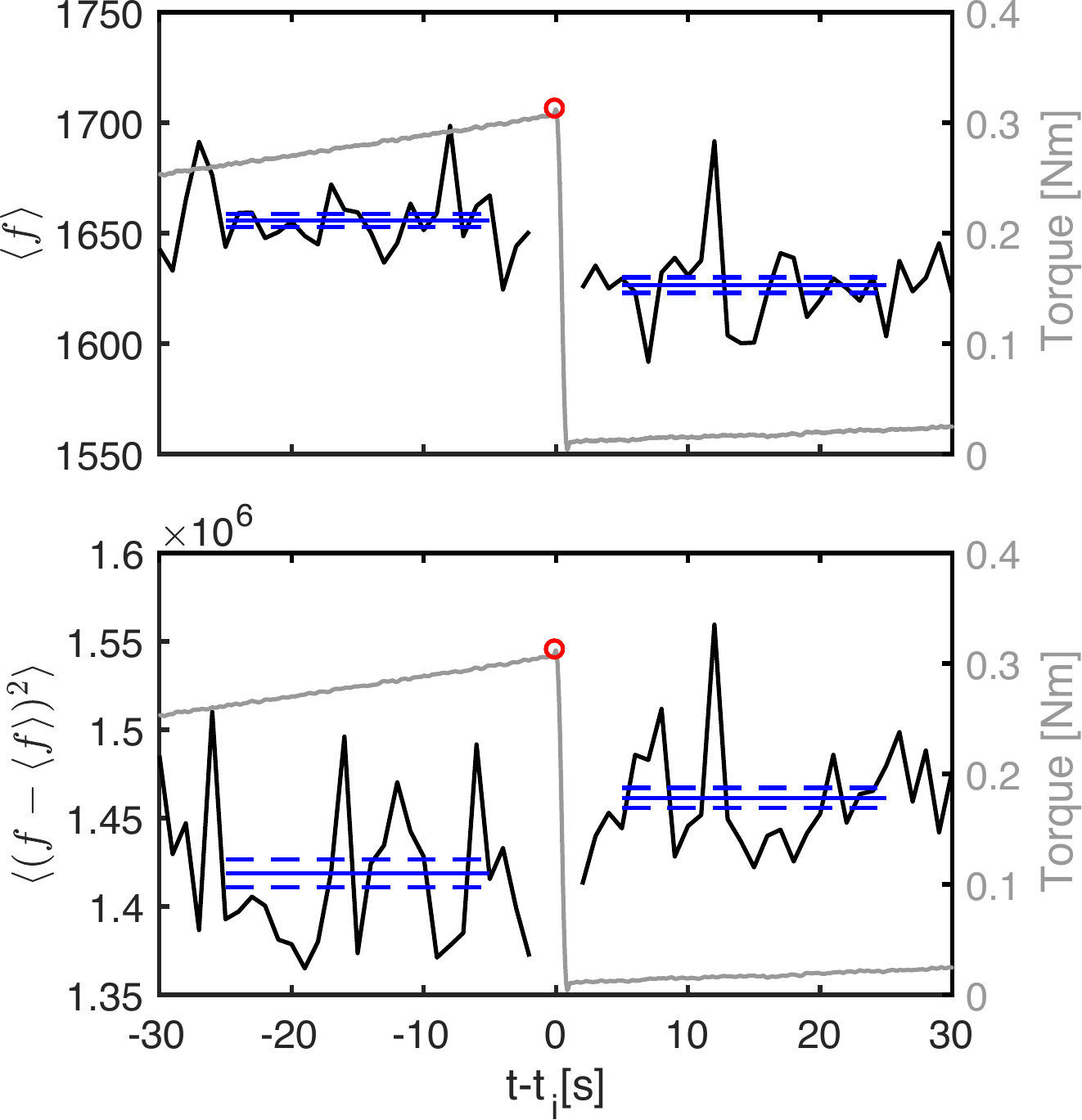}
\vskip.125in
%\begin{table}[]
%\centering
\begin{tabular}{l|ccc}
         & \multicolumn{1}{l}{Significant Rise} & \multicolumn{1}{l}{Significant Drop} & \multicolumn{1}{l}{Null} \\ \hline\hline
mean     & 9\%                                  & {\bf 34\%}                                 & 57\%                     \\
variance      & {\bf 30\% }                               & 8\%                                  & 62\%                     \\
kurtosis & 19\%                                 & {\bf 35\%}                                 & 36\%                    
\end{tabular}
%\end{table}
\caption{\label{fig:P6E900}\label{tab:predict}
{(top)} $\langle f \rangle$ and {(middle)} $\langle(f-\langle f \rangle)^{2} \rangle$ for $\DEM$ for a single slip, measured by a single sensor.
The solid blue lines show the mean values for the first and last $28$~seconds respectively and dashed lines show the  $95\%$-confidence in these means.
In both plots, the torque is plotted on the right axis in light grey, with the slip indicated by a red circle.
{(bottom)} The success rates for which we observed statistically significant discontinuities in the mean, variance and kurtosis in $f$ for $\DEMshort$ as measured by a single sensor and at a time coincident with a slip. Boldface values are those that match the sign of the average changes.% in Fig.~\ref{fig:MeanFreq}.
}
\end{figure}

Finally, we consider $\DEMshort$ as a possible metric for characterizing single-slip, single-sensor measurements.
In Fig.~\ref{fig:P6E900} we examine a characteristic example for a period centered on a single slip. 
$\langle f \rangle$ (top) and $\langle (f -\langle f \rangle)^{2}\rangle$ (middle) of the excited modes, calculated based on a \emph{single} sensor are plotted against $t-t_i$. To highlight the stepwise change at $t=t_i$, we additionally plot the mean$\pm2\sigma$ as horizontal lines during the pre- and post-phases, and plot the torque on the right axis in both plots. 
As for the ensemble-averaged data (Fig.~\ref{fig:MeanFreq}), we observe a significant drop in $\langle f\rangle$ coincident with the slip at $t=t_i$ accompanied by a rise in the variance.
We perform this same analysis for all slips that are well-separated in time, for all 12 piezos, for {$887\times12=10644$} sets of statistics of the type shown in Fig.~\ref{fig:P6E900}. 
The results of this analysis are summarized in the table in Fig.~\ref{fig:P6E900} (bottom).
We find that single sensor measurements are consistent with the trends depicted in Fig.~\ref{fig:MeanFreq}, as emphasized by the boldface in Fig.~\ref{fig:P6E900} (bottom).
Importantly, some slips will produce a detectable signal in these metrics for only one or a subset of sensors. Moreover, some slips produce no signal at any sensor due to the dissipative nature of the system and because the $12$~sensors, each approximately $5$~mm wide, cover less than $1\%$ of the length of the perimeter. 

The analysis we present here represents an important step in connecting passive acoustic measurements directly to the state of the material.
While acoustic emissions have previously been known to coincide with the failure of granular media, our method provides a new capability: assessment of the progress of a system en route to failure.
The shift observed in moments of the occupancy of vibrational modes is consistent with observations that granular systems under less stress exhibit an excess of floppy, low-frequency modes \cite{Owens2013}, and can be connected more broadly to similar observations in jammed solids as the volume fraction is reduced  \cite{Ghosh2010,Silbert2005,Chen2010} and as shear progresses   \cite{Manning2011}.
Our results indicate \IMP{promise for predictive forecasting of failure in slowly sheared, disordered systems}. However, any approach to forecasting of this sort is most likely to be probabilistic rather than deterministic: $\aDEMshort$ may signal an \emph{increased likelihood} for an event, particularly with better sensor coverage where passive measurements may more completely reveal the features of the density of vibrational modes. These techniques also provide a route for \IMP{improved characterization of the vibrational properties of disordered materials.}

Assessing the internal stress state of a granular system is notoriously difficult: photoelastic, optical, and tomographic techniques  \cite{Majmudar2005,Brodu2015,Hurley2016} require specialized materials or slow scanning times to make quantitative measurements of internal stresses.
We have found that the density of excited vibrational modes $\DEMshort(f)$ appears to provide \IMP{a new technique for reporting the internal stress in the system} which does not require optical access, and can be applied to 3D systems with fast measurement times.
Importantly, this method does not require acoustic-driving, which risks triggering a failure in materials near threshold, and should work on a variety of materials.
Furthermore, \IMP{this analysis is largely independent of the details of the sensor mechanism,} so an obvious next step is to test the approach with seismic data.

\begin{acknowledgments} This work was supported by NSF Grant DMR-1206808 and the James S. McDonnell Foundation. We thank Paul Johnson, Craig Maloney, Corey O'Hern and Eli Owens for helpful discussions, and  Cross Automation for technical assistance with our motion control needs.

\end{acknowledgments}

\bibliography{ExcitedModes}

\end{document}